\documentclass[12pt,aps,prd,a4paper,superscriptaddress,nofootinbib,onecolumn]{revtex4}

\usepackage{amsfonts}
\usepackage{amsmath}
\usepackage{babel}
\usepackage[active]{srcltx}
\usepackage{graphicx}
\usepackage{color}
\usepackage{float}
\usepackage{changebar}
\usepackage{esint}
\usepackage{dsfont}
\usepackage{multirow}
\usepackage[usenames,dvipsnames]{xcolor}
\usepackage[colorlinks=true,pdfstartview=FitV,linkcolor=blue,citecolor=blue,urlcolor=blue,breaklinks=true]{hyperref}
\usepackage{braket}
\usepackage{mathtools}
\usepackage{slashed}
\usepackage{ulem}
\usepackage{empheq}
\usepackage{multirow}
\usepackage[utf8]{inputenc}
\usepackage[T1]{fontenc}
\usepackage{amssymb}
\usepackage{subfigure}
\makeatletter
\RequirePackage{color}
\newcommand{\e}{\textrm{e}}

\newcommand{\revision}{\textcolor{black}}

\begin{document}

\title{On the asymmetric non-canonical braneworld in five dimensions}

\author{F. C. E. Lima}
\email{cleiton.estevao@fisica.ufc.br (F. C. E. Lima)}
\affiliation{Programa de P\'{o}s-gradua\c{c}\~{a}o em F\'{i}sica, Universidade Federal do Maranh\~{a}o, Campus Universit\'{a}rio do Bacanga, S\~{a}o Lu\'{i}s, MA, Brazil.}

\author{F. M. Belchior}
\email{belchior@fisica.ufc.br (F. M. Belchior)}
\affiliation{Departamento de F\'{\i}sica, Universidade Federal do Cear\'{a}, Campus do Pici, Fortaleza, CE, Brazil.}

\author{C. A. S. Almeida}
\email{carlos@fisica.ufc.br (C. A. S. Almeida)}
\affiliation{Departamento de F\'{\i}sica, Universidade Federal do Cear\'{a}, Campus do Pici, Fortaleza, CE, Brazil.}
\affiliation{Departamento de F\'{i}sica Te\'{o}rica, Universidad de Valencia, Burjassot, Valencia, Spain.}

\author{P.K. Sahoo}
\email{pksahoo@hyderabad.bits-pilani.ac.in (P.K. Sahoo)}
\affiliation{Department of Mathematics, Birla Institute of Technology and Science-Pilani, Hyderabad Campus, Hyderabad-500078, India.}

\begin{abstract}
\vspace{0.5cm}
\noindent \textbf{Abstract:} Revisiting Einstein's gravitational theory, we build a five-dimensional braneworld. Within this framework, one announces the appearance of symmetric and asymmetric domain walls. Furthermore, it examines the emergent four-dimensional gravity from a theory with non-canonical dynamics. Exploring the physical and mathematical aspects, e.g., brane's energy density and the Kaluza-Klein (KK) spectrum, one verifies that brane splitting is absent in the canonical and non-canonical theories. Additionally, we note the localization of the four-dimensional fluctuation projection on the 3-branes, which ensures the theory's stability. Thereby, one can conclude that the behavior of gravitational perturbations of the domain wall maintains a profile similar to a stable and non-localizable tower of massive modes. In contrast, within the brane core, the matter sector generates new barriers and potential wells, resulting in massive modes with approximately symmetric amplitudes. However, the non-canonical dynamics generate massive modes with asymmetric amplitudes far from the 3-brane.
\end{abstract}
\maketitle

\thispagestyle{empty}

\newpage

\section{Introduction}\label{secI}

Randall and Sundrum (RS), motivated by the evidence that our universe is within a four-dimensional compacted spacetime, proposed a high-dimensional gravity theory \cite{RS1,RS2}. Briefly, this motivation arises due to the premise that stems from the hypothesis that standard model matter cannot propagate over large distances in extra dimensions without conflicting with observable data. However, one can bypass these constraints if the standard model is limited to a four-dimensional subspace \cite{Arkani,Antoniadis,Shiu}. This compact four-dimensional subspace is called brane or 3-brane. RS's first proposal aims to address the hierarchy problem \cite{RS1}. Meanwhile, RS proposes a second prototype, i.e., a 3-brane within a five-dimensional spacetime \cite{RS2}. Thus, one announces that gravity is reproduced more accurately in high-dimensional theories.

After the alluring advancements proposed by RS, Gremm's work emerged, announcing a study of four-dimensional gravity on a thick domain wall \cite{Gremm}. In his manuscript, Gremm employs first-order formalism to investigate the domain walls in AdS space.  Briefly, Gremm conducts this study adopting the action
\begin{align}\label{00}
    S=\int\,d^4x\,dy\, \sqrt{-g}\left[\frac{1}{4}R-\frac{1}{2}(\partial\phi)^2-V(\phi)\right].
\end{align}
Although simple, Gremm's proposal stands out for the elegance of its structure, which resembles a relativistic quantum theory. This feature enables the analysis of the graviton dynamics through a Schr\"{o}dinger-like theory. Thus, a question arises regarding Gremm's proposal, i.e., promoting Gremm's model to a non-canonical theory: How do non-canonical dynamics affect thick domain walls? We will investigate this issue in this manuscript.

Closely related to this discussion, the search for a theory that would unify gravitation and electromagnetism using five-dimensional Riemannian geometry and its generalizations has become a focus for particle physicists \cite{Kerner,Cho1,Cho2}. \revision{This search leads us to the development of the Kaluza-Klein (KK) theory \cite{Kaluza,Klein,Scherk}.} 
However, with the emergence of theories involving extra dimensions, e.g., supergravity \cite{Cremer1,Cremer2} and later braneworld theory \cite{RS1,RS2}, the KK theory acquired new purposes. In the context of supergravity, the KK theory focuses on detecting compactified extra dimensions \cite{Bailin}. Concurrently, in the braneworld framework, the KK theory provides insights into the graviton's bound states in higher dimensions \cite{RS2}. Thus, one can affirm that the KK theory describes gravitational fluctuations in higher dimensions in terms of the four-dimensional KK states, characterized by mass eigenvalues $m^2$ \cite{Belchior1,Belchior2}. Once we know the importance of the KK theory, another issue arises, i.e., what is the effect of the non-canonical dynamic in the KK spectrum of a five-dimensional braneworld?

In this paper, we will discuss a theory with a non-canonical dynamic. In other words, we are substituting the usual dynamics for $\frac{f(\phi)}{2}\partial_\mu\phi\partial^\mu\phi$. Indeed, one applies this approach in several studies seeking to deform domain walls and examine new classes of field solutions \cite{FLima}. Initially, Lee and Nam were pioneers in studies of non-canonical theories \cite{Lee}. However, in his work, they imposed the existence of a generalizing function (i.e., a dielectric function) in the gauge sector. In the gravitational framework, the motivation for studying theories with non-canonical dynamics again appears from the search for explanations of inflationary evolution \cite{Picon1}. In this scenario, the absence of interactions may lead us to inflationary evolution theories due to the non-canonical term \cite{Picon2,Picon3}. Thereby, while in topological models, one expects that non-canonical dynamics provide new structures, in the gravitational scenario, we believed that non-canonical theory could explain the universe's accelerated evolution. In our work, we seek to understand the influence of the non-canonical term in thick brane theory.

Indeed, we believe that the non-canonical terms may give rise to a theoretical formulation of an asymmetric braneworld. Therefore, it is essential to highlight that asymmetric braneworld models have been the subject of intense study \cite{Gergely,ZLin,Roldao,Rosa}. In particular, asymmetric braneworld configurations draw significant attention due to the innovative results emerging in this framework. Among the wide range of asymmetric theories in the braneworld framework, one highlights some studies of notable importance. For instance, Takahashi and Shiromizu announced that considering the braneworld devoid of the $\mathds{Z}_2$ symmetry with long wave approach, the braneworld asymmetry appears naturally \cite{Takahashi}. Also, one finds an asymmetric hybrid braneworld using the scalar field as the source in Ref. \cite{Menezes}. Additionally, motivated by the quantum gravity approach, i.e., adopting the path integral formalism, solutions of asymmetric canonical instantons have been reported \cite{HZhang}. Hence, these investigations motivate us to seek to identify a new class of asymmetric branes by manipulating the generalizing function $f(\phi)$.

We structure the manuscript into four sections. In section \ref{secII}, one presents the general framework of the theory. Furthermore, we adopt the warped geometry to study a canonical and non-canonical (exponential) braneworld. Posteriorly, i.e., in section \ref{secIII}, one examines the brane's stability, massless modes (zero modes), and massive modes. Finally, one announces our findings in section \ref{secIV}.

\section{General framework}\label{secII}

The non-canonical theories have sparked increasing interest \cite{Casana1,Casana2}. For instance, these theories are typically employed in flat spacetimes to deform structures and generate field configurations with internal structures or to compactify the field profiles \cite{Lima1,Lima2}. Therefore, one concludes that this mechanism is advantageous in dealing with physical problems at flat spacetime once it is responsible for geometrically deforming domain walls into double domain walls. Mindful of these peculiarities, we will examine the effects that a non-canonical theory generates in a higher-dimensional theory, specifically a five-dimensional braneworld. To accomplish our purpose, let us consider the action 
\begin{align}\label{Eq1}
    S=\int\, d^4x dy\, \sqrt{-g} \bigg[\frac{1}{4}R-\frac{1}{2}f(\phi)\partial_M\phi\partial^M\phi-V(\phi)\bigg].
\end{align}
In this framework, $f(\phi)$ is the generalizing function. This function is responsible for producing a non-canonical theory. Note that if $f(\phi)\to 1$, the theory recovers the standard theory as announced by Gremm \cite{Gremm}. Additionally, $g$ is the determinant of the background metric, $\phi$ is the matter field, $R$ represents the five-dimensional scalar curvature, and $V(\phi)$ is the potential.

Let us now examine the equations of motion. To fulfill this, we vary the action concerning the metric and the matter field, respectively. Thereby, Einstein's equation is 
\begin{align}\label{Eq2}
    R_{MN}-\frac{1}{2}R g_{MN}=2T_{MN},
\end{align}
and the equation for the scalar field is 
\begin{align}\label{Eq3}
    f\partial_M\partial^M\phi+\frac{1}{2}f_\phi\partial_M\phi\partial^M\phi=V_\phi.
\end{align}
Here, $R_{MN}$ is Ricci's tensor, and $T_{MN}$ is the stress-energy tensor. Explicitly, the stress-energy tensor is
\begin{align}\label{Eq4}
    T_{MN}=f\partial_M\phi\partial_N\phi+g_{MN}\mathcal{L}_\textrm{matter},
\end{align}
where $\mathcal{L}_\textrm{matter}$ is the Lagrangian density related the matter.

\subsection{Description of the spacetime metric}\label{secIIa}

In this study, one employs a warped geometric background which preserves the four-dimensional Poincaré invariance with the extra-dimensional coordinate extending to infinity. Therefore, assuming a five-dimensional bulk, the spacetime metric \cite{Gremm} is
\begin{align}\label{Eq5}
ds^2_5=\e^{2A(y)}\eta_{\mu\nu}dx^\mu dx^\nu+dy^2, 
\end{align}
where $A(y)$ represents the warp factor. Consequently, $\textrm{e}^{2A(y)}$ is the warp factor. Additionally, $\eta_{\mu\nu}=$ diag$(-1,+1,+1,+1)$, i.e., it describes conventional Minkowski spacetime where $\mu,\nu=0,1,2,3$. Furthermore, one highlights that Latin indices are $M, N=0,1,2,3,4$.

\subsection{The five-dimensional braneworld}\label{secIIb}

For the proposed conjecture, we consider a static braneworld coupled to a real scalar field $\phi$, whose action is specified in Eq. \eqref{Eq1}. In this context, one assumes that the scalar field is strictly a function of the extra-dimensional coordinate, i.e., $\phi\equiv\phi(y)$. Considering this premise, Eqs. [\eqref{Eq2}-\eqref{Eq3}] yield the equations:
\begin{align}\label{Eq6}
    &f\phi''+4fA'\phi'+\frac{1}{2}f_\phi \phi'^2=V_\phi;\\ \label{Eq7}
    &A''=-\frac{2}{3}f\phi'^2;\\ \label{Eq8}
    &A'^2=\frac{1}{6}f\phi'^2-\frac{1}{3}V.
\end{align}
To solve the system of equations [\eqref{Eq6}-\eqref{Eq8}], we employ the approach known as the first-order formalism. 
In summary, the first-order formalism involves an \textit{ansatz} that reduces the order of the equations of motion without saturating the theory's energy. This approach was initially motivated by Gremm \cite{Gremm}. In his work, Gremm used the first-order formalism motivated by applications of the BPS supergravity theory, as proposed in Refs. \cite{DeWolfe,Cvetic,Skenderis}. Thereby, using an approach similar to the proposed in Ref. \cite{Gremm}, one obtains the non-canonical first-order equations, viz,
\begin{align}\label{Eq9}
    &\phi'=\frac{1}{2f}W_\phi,\\ \label{Eq10}
    &A'=-\frac{1}{3}W,\\ \label{Eq11}
    &V=\frac{1}{8f}W_\phi^2-\frac{1}{3}W^2.
\end{align}
Here, $W\equiv W(\phi)$ is the superpotential, i.e., an auxiliary function implemented to help us to reduce the order of Eqs. [\eqref{Eq6}-\eqref{Eq8}]. Therefore, once we know the form of the function $W(\phi)$, it is possible to solve the equation's system of the five-dimensional braneworld.

Let us first examine the brane's energy before delving into the study of its equation of motion. In this case, one obtains the energy density considering the first component of the stress-energy tensor, i.e., $T_{00}$. Thus, the brane's energy density is
\begin{align}\label{Eq12}
    \rho(y)=T_{00}=-\textrm{e}^{2A}\mathcal{L}_{\textrm{matter}},
\end{align}
which leads us to 
\begin{align}\label{Eq13}
    \rho(y)=\frac{1}{2}\frac{d}{dy}(\textrm{e}^{2A}W).
\end{align}

Therefore, the brane's energy is 
\begin{align}\label{REq13}
    \textrm{E}=\frac{1}{2}\textrm{e}^{2A(y\to\infty)}[W(\phi_\infty)-W(\phi_{-\infty})].
\end{align}
By analyzing the energy \eqref{REq13}, one notes that the energy of the brane assumes a constant value, i.e., a zero value. Therefore, we can assert that the total energy of the first-order solutions converges to a finite value, corresponding to the Bogomol'nyi bound.

To continue, we need to adopt a specific form for the superpotential. Notably, one selects a superpotential that engenders a theory with $\mathds{Z}_2$ symmetry in the matter sector, resulting in the appearance of domain walls \cite{DeWolfe}. Thereby, we consider the polynomial superpotential 
\begin{align}\label{Eq15}
  W(\phi)=2\lambda \left(\nu^2 \phi-\frac{\phi^3}{3} \right).
\end{align}
Finally, assuming the first-order equations, the brane's energy density, and the superpotential, we are ready to examine the five-dimensional braneworld. One performs this study in the subsequent subsections.

\subsubsection{The standard case}

To analysis the effects of the non-canonical theory, it is necessary to understand the results stemming from the canonical theory, i.e., $f\to 1$. Thus, we start by investigating the standard (or canonical) case. In this framework, the equations of motion [\eqref{Eq6}-\eqref{Eq8}] takes the form
\begin{align}\label{Eq16}
    &\phi''+4A'\phi'=V_\phi;\\ \label{Eq17}
    &A''=-\frac{2}{3}\phi'^2;\\ \label{Eq18}
    &A'^2=\frac{1}{6}\phi'^2-\frac{1}{3}V.
\end{align}

\begin{figure}[!ht]
  \centering
  \includegraphics[height=6cm,width=7cm]{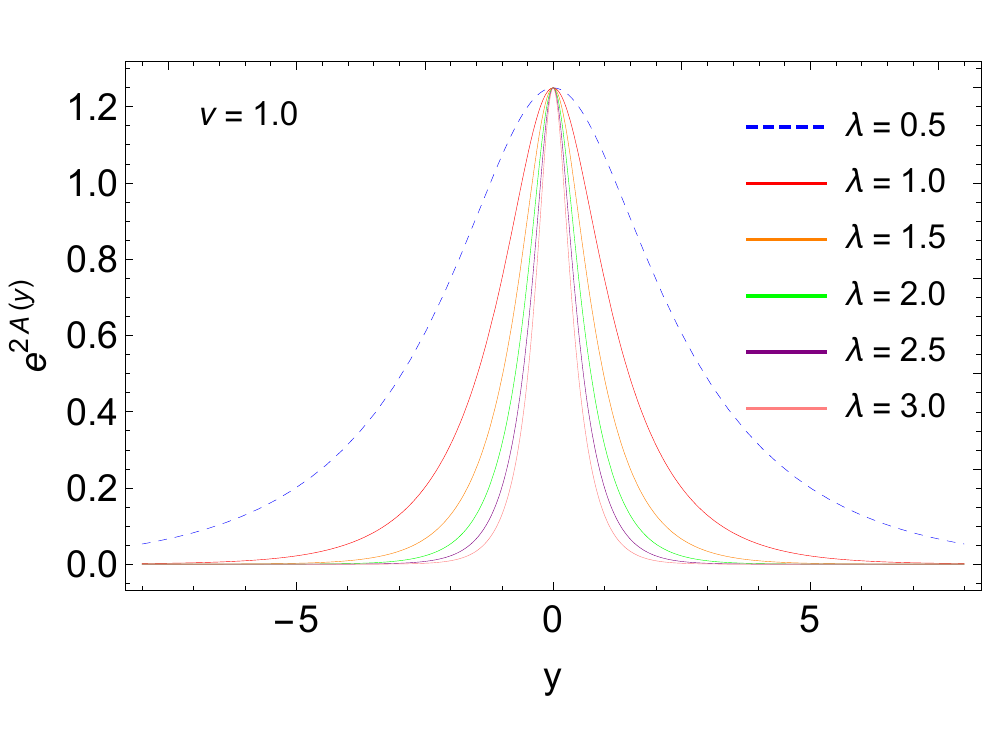}\vspace{-1.2cm}
  \caption{Warp factor vs. extra-dimensional coordinate.}  \label{fig1}
\end{figure}

\begin{figure}[!ht]
\caption{Analytical solutions of the five-dimensional braneworld.}  \label{fig2}
  \centering
  \subfigure[Scalar field vs. extra-dimensional coordinate.]{\includegraphics[height=4.3cm,width=5.3cm]{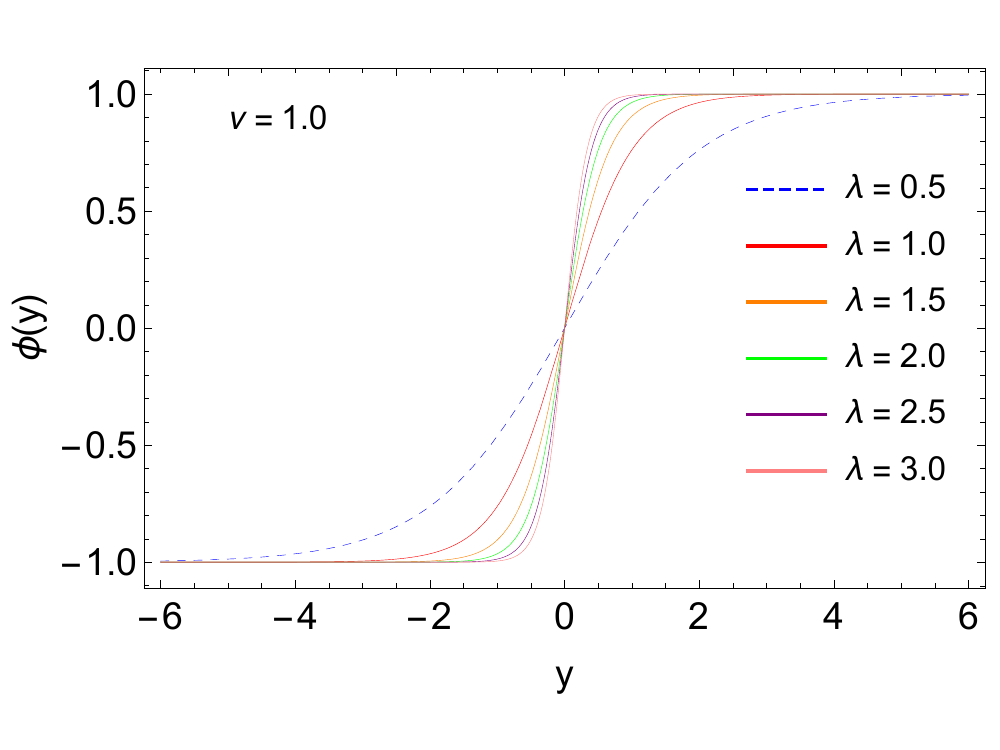}}\hfill
  \subfigure[Potential vs. matter field.]{\includegraphics[height=4.3cm,width=5.3cm]{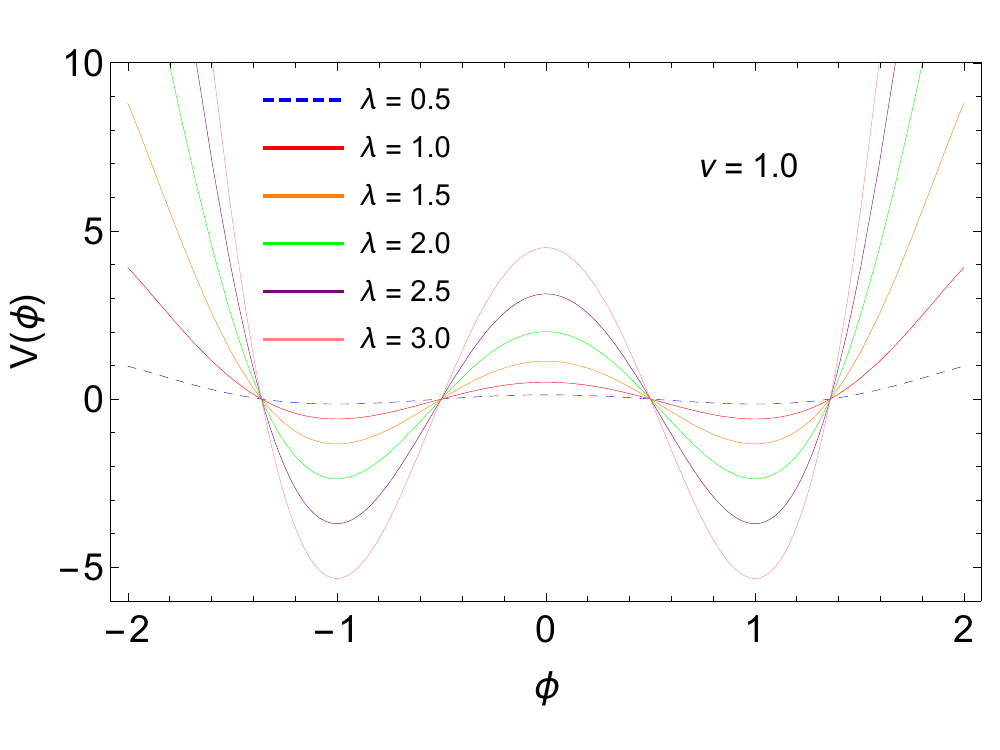}}\hfill
  \subfigure[Brane's energy density vs. extra-dimensional coordinate.]{\includegraphics[height=4.3cm,width=5.3cm]{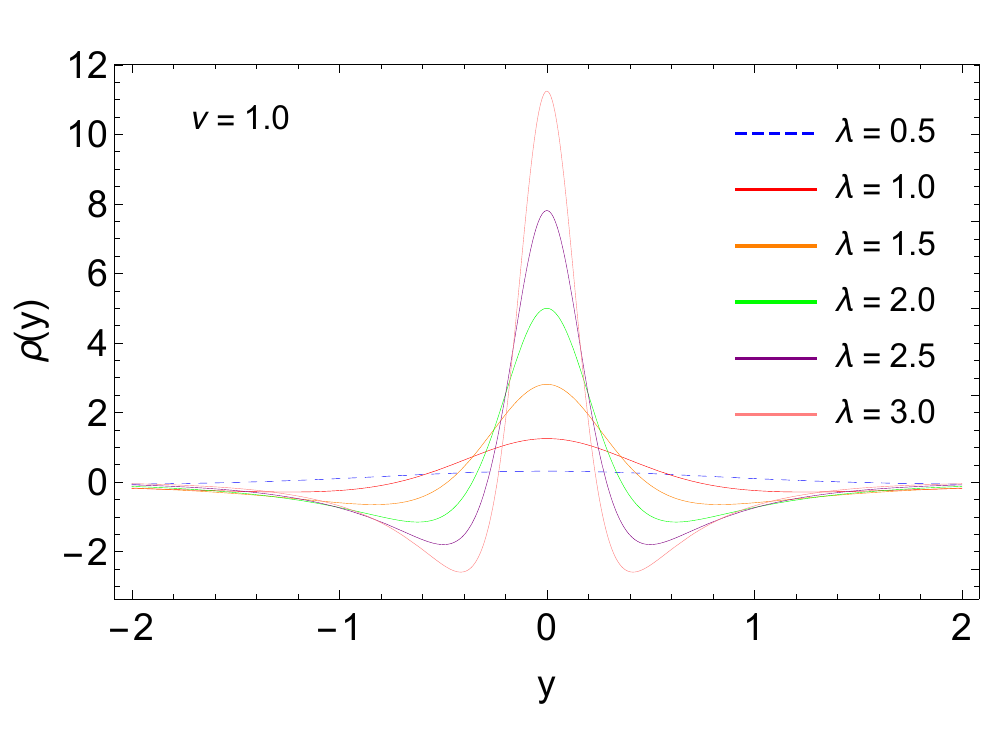}}
\end{figure}

Now, let us examine the brane's solutions [\eqref{Eq16}-\eqref{Eq18}]. In this framework, the solutions for the matter field, warp function, and potential are
\begin{align}\label{Eq19}
    &\phi(y)=\nu\tanh(\mu\, y),\\ \label{Eq20}
    &A(y)=\frac{\nu^2(2\textrm{ln}\,[\textrm{sech}\,(\mu y)^2]+ \textrm{sech}\,(\mu y)^2)}{9},\\ \label{Eq21}
    &V(\phi)=\frac{1}{6}\lambda^2 \bigg[3(\phi^2-\nu^2)^2-\frac{8}{9}\bigg(\phi^3-3\nu^2\phi\bigg)^2\bigg],
\end{align}
where $\mu=\lambda\nu$. The analytical results of Eqs. [\eqref{Eq19}-\eqref{Eq21}] are presented in Figs. \ref{fig1} and \ref{fig2}. These findings demonstrate that the five-dimensional braneworld is a thick brane theory. Additionally, one changes the brane's width by varying the parameter $\lambda$. In this framework, the thick warp geometry is responsible for the spontaneous symmetry breaking in the matter sector. Thereby, one has a theory with symmetry breaking featuring two symmetric energy potential minima, i.e., a Higgs-like potential. The emergence of this potential induces the appearance of single-domain walls. In this case, the domain walls are true kinks localized around the 3-brane. We confirm the localization of domain walls on the 3-brane by the symmetry and localization of the brane's energy density at $y=0$.

\subsubsection{The asymmetric case}

Let us direct our attention to the study of a non-canonical braneworld. To investigate the non-canonical case, allow us to select a specific generalizing function. In this paper, we will adopt an exponential generalizing profile, viz.,
\begin{align}\label{Eq22}
    f(\phi)=\textrm{exp}\left(\frac{p\,\phi}{\nu}\right) \hspace{0.5cm} \textrm{with} \hspace{0.5cm} p\in\mathbb{N}^{*}.
\end{align}
We assume this generalization because this generalizing function can result in a geometrical deformation along the extra-dimensional coordinate. Thereby, one can obtain a new class of braneworld \cite{Matheus}. In Ref. \cite{Lima3}, models with non-polynomial generalizing functions inherently alter the form of the topological structure of the matter field, so modifying the brane's physical structures \cite{Lima3}. Additionally, one highlights that exponential generalizing functions have been widely employed in cosmological scenarios to investigate inflationary models \cite{Santos}. Therefore, these hypotheses and applications help us motivate the choice of the generalizing function \eqref{Eq22}.

\begin{figure}[!ht]
  \centering
  \includegraphics[height=7cm,width=8cm]{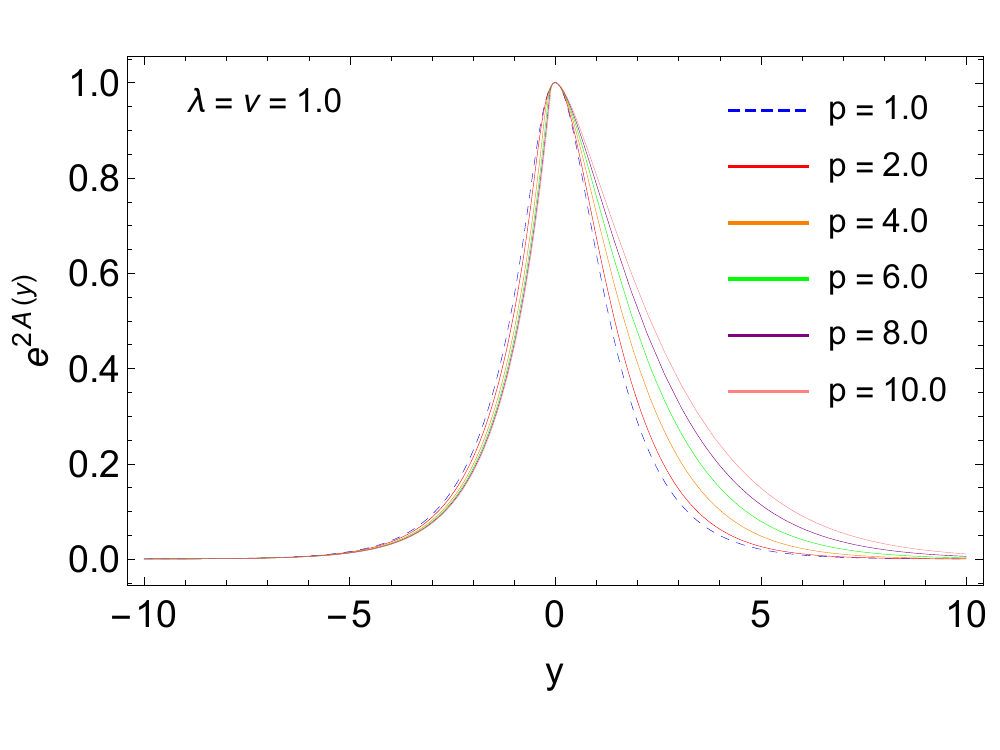}\vspace{-1.2cm}
  \caption{Warp factor vs. extra-dimensional coordinate.}  \label{fig3}
\end{figure}

\begin{figure}[!ht]
\caption{Numerical solutions of the five-dimensional braneworld.}  \label{fig4}
  \centering
  \subfigure[Scalar field vs. extra-dimensional coordinate.]{\includegraphics[height=4.3cm,width=5.3cm]{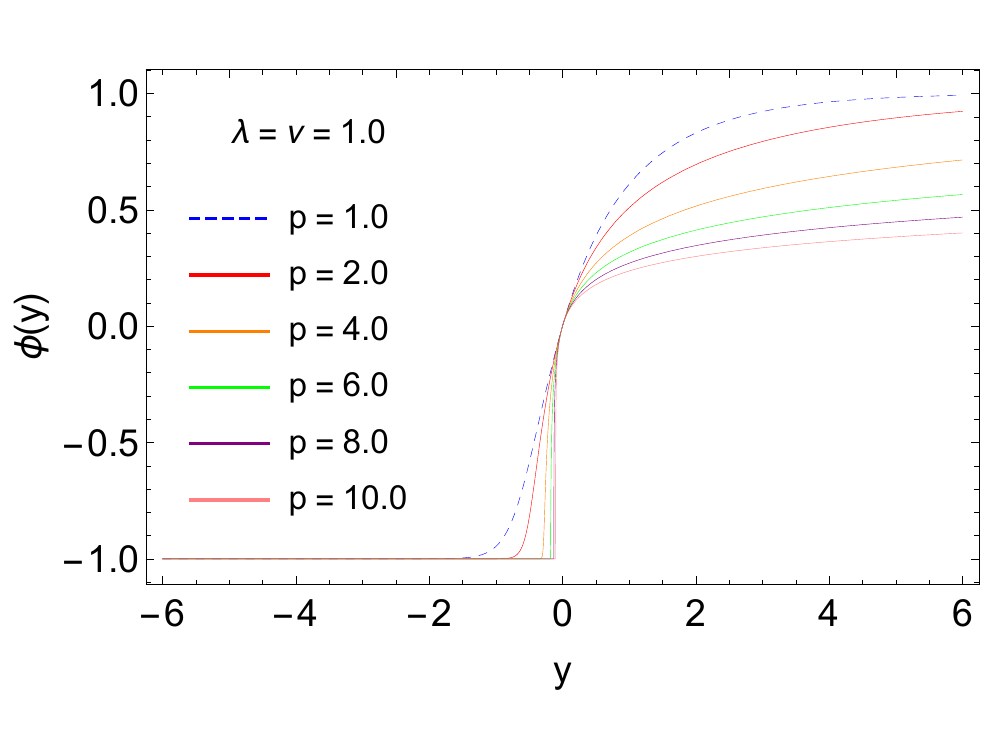}}\hfill
  \subfigure[Potential vs. matter field.]{\includegraphics[height=4.3cm,width=5.3cm]{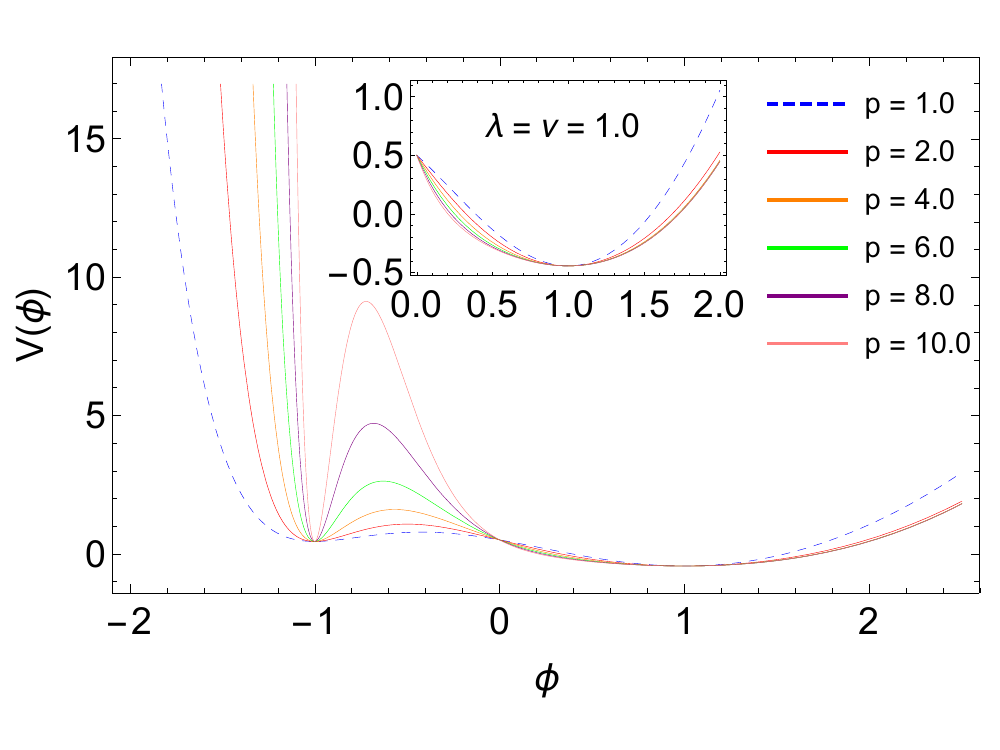}}\hfill
  \subfigure[Brane's energy density vs. extra-dimensional coordinate.]{\includegraphics[height=4.3cm,width=5.3cm]{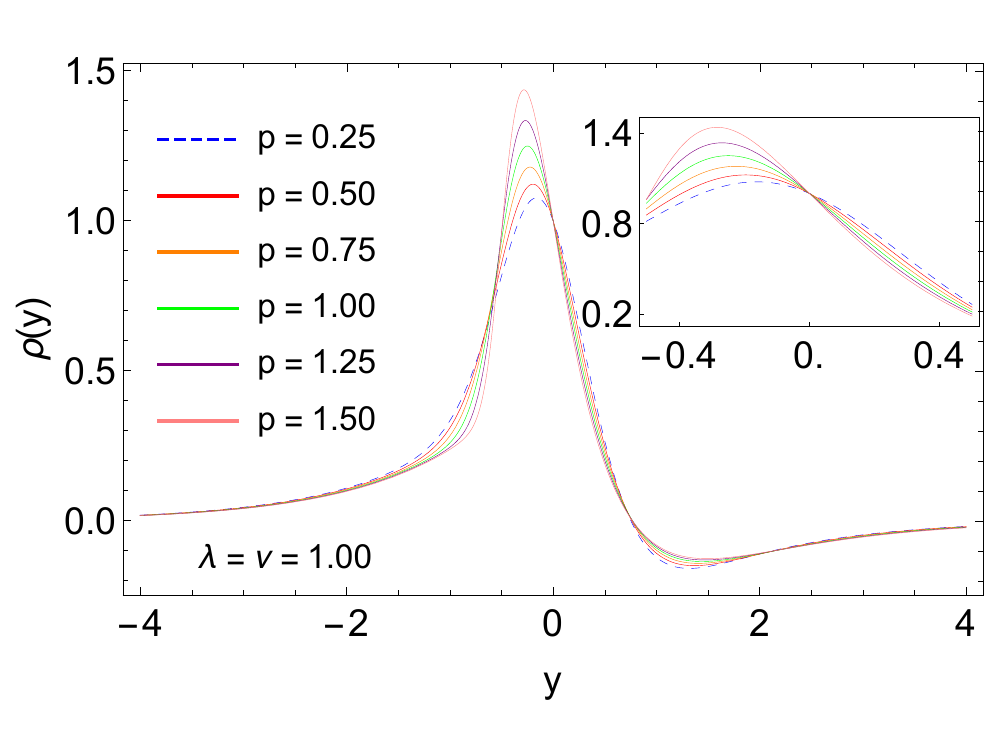}}
\end{figure}

Allow us to reformulate the equations of motion [\eqref{Eq9}-\eqref{Eq11}] adopting the generalized function \eqref{Eq22}. Thus, one obtains
\begin{align}\label{Eq23}
    &\phi'=\frac{1}{2}\lambda\text{e}^{-\frac{p\phi}{\nu}}(\nu^2-\phi^2),\\ \label{Eq24}
    &A'=-\frac{2}{3}\lambda\left(\nu^2\phi-\frac{\phi^3}{3}\right),\\ \label{Eq25}
    &V=\frac{1}{6}\lambda^2\left[\textrm{e}^{-\frac{p\phi}{\nu}}(\nu^2-\phi^2)^2-\frac{8}{9}(\phi^3-3\nu^2\phi)^2\right].
\end{align}
Upon inspecting the system of equations [\eqref{Eq23} to \eqref{Eq25}], the need arises to employ a numerical method to analyze the non-canonical braneworld solutions. Therefore, to proceed, let us utilize the numerical interpolation method. Fundamentally, the interpolation method involves discretizing the independent variable domain into $m$ points, i.e., $y_0,\, y_1,\, y_2,\, \dots,\,  y_m$. Discretizing the independent variable domain, one can estimate the solutions for $A(y)$, $\phi(y)$, and $V(\phi)$ at intermediate points\footnote{One can find more details on the interpolation method in Ref. \cite{Hildebrand}}. We present the numerical results of the non-canonical braneworld with the exponential generalizing function in Figs. \ref{fig3} and \ref{fig4}.

The numerical solutions exposed in Figs. \ref{fig3} and \ref{fig4} suggest the existence of a thick five-dimensional asymmetric braneworld. In this conjecture, the warped geometry of the thick brane becomes asymmetric by varying the $p$-parameter. Consequently, one notes that the asymmetry from the spacetime infers in the matter sector, which leads to the emergence of vacuum decay. This phenomenon refers to a hypothetical transition from one vacuum state to another with lower energy. Thus, the potential profile induces a geometric deformation of the matter field, i.e., a deformation of the domain wall, making the emitted matter field from the 3-brane more compacted at $y \in (-\infty, 0]$. Finally, one concludes that the non-canonical thick brane is asymmetric. This asymmetry is confirmed by analyzing the brane's energy density\footnote{We note numerically that the asymmetric brane has null energy due to the asymptotic profile of the warp function (i.e., $\textrm{e}^{2A(y_\infty)}\to 0$), which leads us to the results correspond to the Bogomol'nyi bound.}.

\begin{center}
    \textit{The asymptotic solutions of the asymmetric brane}
\end{center}

Let us analyze the asymptotic behavior of the matter field and the warp factor. We will begin by examining the behavior of these entities on the brane, i.e., at $y=0$. To fulfill our aim, we will utilize Eqs. \eqref{Eq23}-\eqref{Eq25}. Hence, it follows that
\begin{align}\label{R1}
    \phi(y)\approx \nu\left(1-\frac{1}{p}\textrm{e}^{p\lambda \nu y}\right)+\dots,
\end{align}
and 
\begin{align}\label{R2}
    A(y)\approx A_0 - \frac{2\lambda \nu}{9p^3}\left[-\frac{3\nu}{2\lambda}\textrm{e}^{2p\lambda\nu y}+\frac{\nu}{3\lambda p}\textrm{e}^{3p\lambda\nu y}+\frac{3p(\nu^2-1)}{\lambda\nu}\textrm{e}^{p \lambda\nu y}-p^3y(\nu^2-3)\right]+\dots.
\end{align}

Meanwhile, for $y\to \pm\infty$, one arrives at
\begin{align}\label{R3}
    \phi(y)\approx \pm\nu+\nu\textrm{e}^{\mp\lambda\nu y\, \textrm{e}^{\mp p}}+\dots
\end{align}
and
\begin{align}\label{R4}
    A(y)\approx A_0 -\frac{4}{9}\lambda\nu^3 y+\dots.
\end{align}

The analysis of the asymptotic behavior of the brane reveals that, near the brane, the solutions related to the matter field and the warp factor change with variations in the generalizing parameter ($p$). Furthermore, one notes that, far away from the brane, i.e., as $p \to \infty$, the generalizing parameter $p$ modifies the asymptotic solutions of the matter field (increasing the matter asymptotic sector in $\nu$, for large $p$-values). Meantime, the warp factor profile remains unchanged. On the other hand, as $y \to -\infty$, this increase becomes negligible when $p$ is large.


\section{The braneworld stability}\label{secIII}

In this section, let us investigate the stability of the five-dimensional braneworld. For this, one considers small perturbations of the metric \eqref{Eq5} in the tensor sector. We perform this analysis due to their relation to the graviton dynamic \cite{Gauy:2022yxj,Xu:2022xxd}.

\subsection{Tensor perturbations}

We start analyzing the braneworld's stability by adopting small tensor perturbations. In this instance, one writes the metric as
\begin{align}\label{metric2g}
ds^2= \textrm{e}^{2A(y)}[\eta_{\mu\nu}+h_{\mu\nu}(x^\mu,y)]dx^{\mu}dx^{\nu}+dy^2.   
\end{align}
Here, $h_{\mu\nu}$ represents the graviton, $x^\mu$ is the position's four-vector. Furthermore, one assumes the $h_{\mu\nu}$ satisfies transverse-traceless (TT) condition, i.e.,  $\partial^\mu h_{\mu\nu}=0$ and $\eta^{\mu\nu}h_{\mu\nu}=0$. Therefore, the perturbed metric is $\overline{g}_{MN}=g_{MN}+g^{(1)}_{MN}$. Explicitly, $g_{MN}$ and $g^{(1)}_{MN}$ are
\begin{align}
g_{MN}=\begin{pmatrix} \textrm{e}^{2A}\eta_{\mu\nu} & 0\\ 0 & 1\end{pmatrix} \hspace{0.5cm} \textrm{and} \hspace{0.5cm} g^{(1)}_{MN}=\begin{pmatrix}
\textrm{e}^{2A}h_{\mu\nu} & 0\\ 0 & 0 \end{pmatrix}.
\end{align}

In this background, the perturbed Ricci tensor is
\begin{align}\label{prt}
R^{(1)}_{\mu\nu}=-\frac{1}{2}\square h_{\mu\nu}-\textrm{e}^{2A(y)}\Big[\frac{1}{2}h^{\prime\prime}_{\mu\nu}+2 A^{\prime} h^{\prime}_{\mu\nu}+(A^{\prime\prime}+4A^{\prime 2})h_{\mu\nu}\Big].    
\end{align}
Using the Eq. \eqref{prt}, one writes perturbed Einstein equation, namely,
\begin{align}\label{pe}
    R^{(1)}_{MN}-\frac{1}{2}g^{(1)}_{MN}R=2T^{(1)}_{MN}.
\end{align}
Meanwhile, by considering small scalar field perturbation ($\vert\epsilon\vert\ll 1$ and $\vert h_{\mu\nu}\vert\ll 1$), i.e., $\phi(y)\rightarrow\overline{\phi}(y)+\epsilon(x^\mu,y)$, we obtain the perturbed stress-energy tensor, viz.,
\begin{align}\label{emt} 
    T^{(1)}\, _{\mu\nu}=-\textrm{e}^{2A(y)}\bigg(\frac{1}{2}f\vert_\epsilon\,\overline{\phi}^{\,\prime\, 2}h_{\mu\nu}+f\vert_\epsilon\,\overline{\phi}^{\,\prime}\epsilon^{\prime}\eta_{\mu\nu}+V h_{\mu\nu}+V_{\phi}\epsilon\eta_{\mu\nu}\bigg).
\end{align}

By substituting Eqs. \eqref{prt} and \eqref{emt} into \eqref{pe}, one obtains the expression for $h_{\mu\nu}$, i.e.,
\begin{align}\label{779g}
    h_{\mu\nu}^{\prime\prime}+4A^{\prime}h_{\mu\nu}^{\prime}=\textrm{e}^{-2A(y)}\square h_{\mu\nu}. 
\end{align}

The Kaluza-Klein (KK) decomposition
\begin{align}\label{777g}
    h_{\mu\nu}(x^\lambda,y)=\sum \widehat{h}_{\mu\nu}(x^\lambda)\chi(y), 
\end{align}
leads us the reformulate Eq. \eqref{779g} as
\begin{align}\label{111g}
\chi^{\prime\prime}(y)+4A^{\prime}\chi^{\prime}(y)=-m^2\textrm{e}^{-2A(y)}\chi(y).
\end{align}

Examining the Eq. \eqref{111g}, it is advisable to remove the term $\textrm{e}^{-2A}$. Thus, we will apply the conformal transformation $dz = \textrm{e}^{-A} dy$ to Eq. \eqref{111g}. This transformation enables us to obtain
\begin{align}\label{113g}
\ddot{\chi}(z)+3\dot{A}\dot{\chi}(z)=-m^2\chi(z).
\end{align}
The dot notation is the derivative concerning the conformal coordinate.

Seeking to write Eq. \eqref{113g} into a Schrödinger-like equation, one considers the transformation  $\chi(z)=\text{e}^{K}(z)\psi(z)$ with $K=-\frac{3}{2}\int\dot{A}\, dz$, which leads us to
\begin{align}\label{114g}
-\ddot{\psi}+V_{\textrm{eff}}\,\psi=m^2\psi,
\end{align}
where $V_{\textrm{eff}}$ is the effective potential, i.e.,
\begin{align}
V_{\textrm{eff}}=\frac{9}{4}\dot{A}^2+\frac{3}{2}\ddot{A}.
\end{align}

\begin{figure}[!ht]
\caption{\revision{Effective potential vs. extra-dimensional coordinate.}}  \label{fig5}
  \centering
  \subfigure[Canonical case.]{\includegraphics[height=7cm,width=8cm]{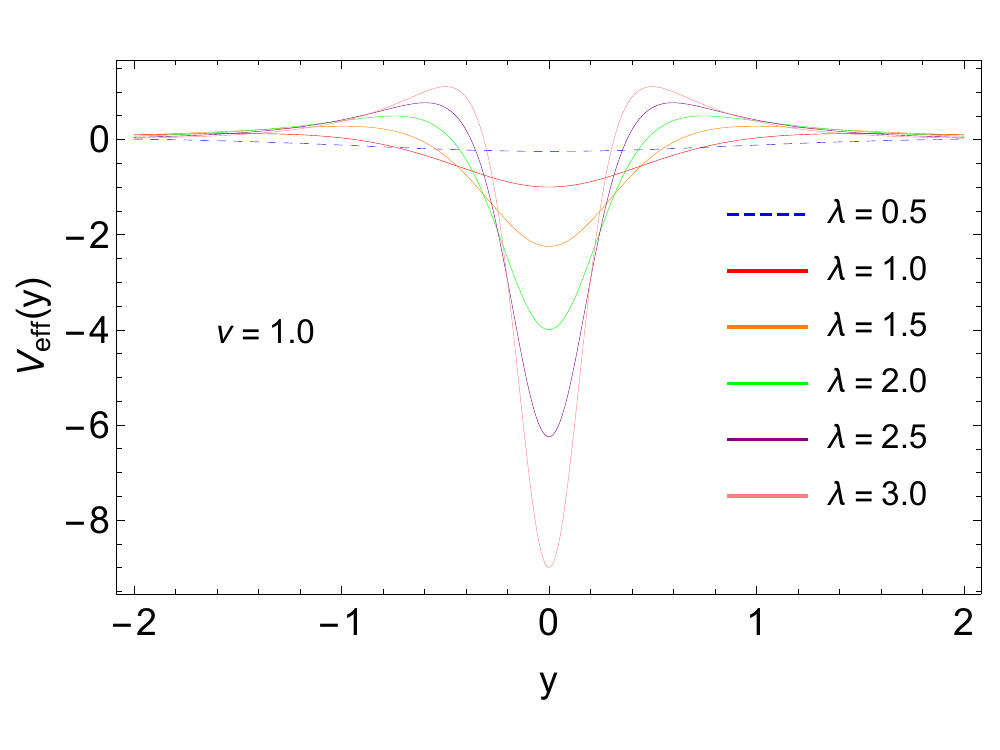}}\hfill
  \subfigure[Non-canonical case.]{\includegraphics[height=7cm,width=8cm]{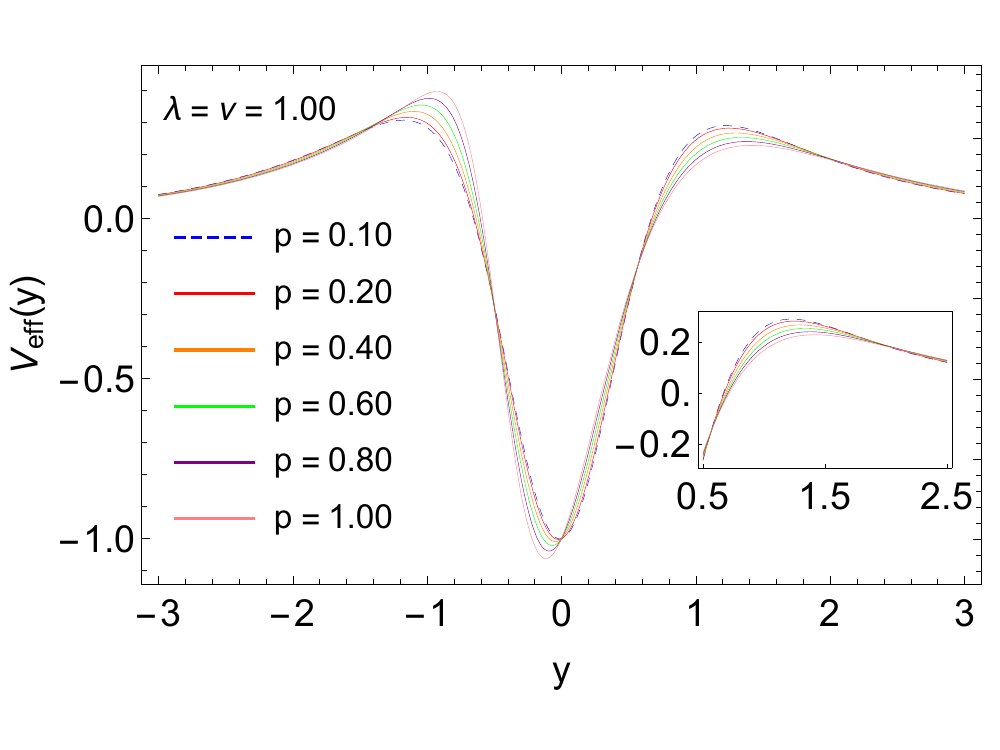}}
\end{figure}

In terms of the extra-dimensional coordinate, the effective potential is
\begin{align}
V_{\textrm{eff}}(y)=\e^{2A}\bigg(\frac{15}{4}A'^2+\frac{3}{2}A'' \bigg).   
\end{align}
One exposes the effective potential in Fig. \ref{fig5}.

The equation \eqref{114g} can be as 
\begin{align}\label{QEq}
    Q^{\dagger}Q\psi=m^2\psi,
\end{align}
with the operators $Q^\dagger$ and $Q$ given by
\begin{align}
    Q^\dagger=\partial_z+\frac{3}{2}\dot{A} \hspace{0.5cm} \textrm{and} \hspace{0.5cm} Q=-\partial_z+\frac{3}{2}\dot{A}.
\end{align}
This factorization implies $m^2\geq 0$, meaning the stability equation supports only states with non-negative eigenvalues. Thus, the braneworld is stable under small metric perturbations.

\begin{figure}[!ht]
\caption{Zero modes vs. extra-dimensional coordinate.}  \label{fig6}
  \centering
  \subfigure[Canonical case.]{\includegraphics[height=7cm,width=8cm]{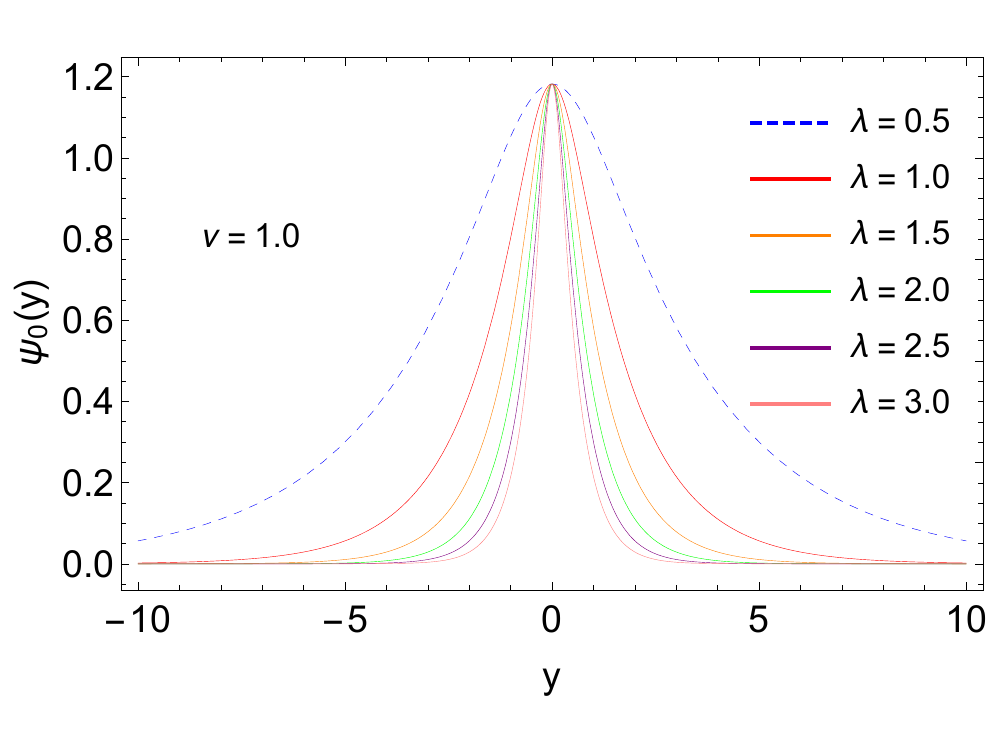}}\hfill
  \subfigure[Non-canonical case.]{\includegraphics[height=7cm,width=8cm]{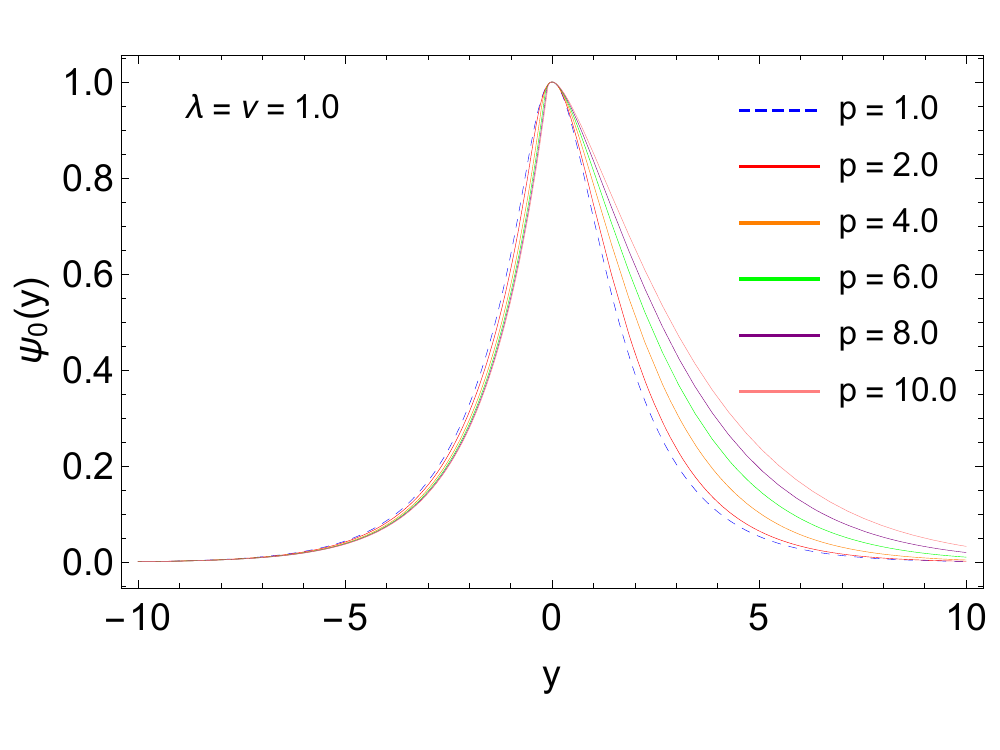}}
\end{figure}

Finally, the graviton zero mode ($m^2=0$) using
\begin{align}
    Q\psi=0,
\end{align}
 which results in
\begin{align}
\psi_0 (y)=\mathcal{N}_0\, \textrm{e}^{\frac{3}{2}A(y)},   
\end{align}
where $\mathcal{N}_0$ is a normalization factor. We display the result of the graviton zero mode in Fig. \ref{fig6}.

Finally, note that the localized four-dimensional gravity then requires zero modes normalizable, i.e.,
\begin{align} 
    \int_{-\infty}^{\infty}dy\, \textrm{e}^{2A(y)}=\int_{-\infty}^{\infty}dz\, \textrm{e}^{3A(z)}<\infty.
\end{align}
Therefore, this condition implies that 
\begin{align}
    \lim_{y\to\pm\infty}\textrm{e}^{2A(y)}=0.
\end{align}

\subsubsection{The massive modes}

Up to this point, one has analytically studied the stability of the KK spectrum. Now, allow us to explore the effects of canonical and non-canonical theory on the massive KK modes. To perform this study, we consider the brane solutions announced in section (\ref{secIIb}) and investigate the solutions of Eq. \eqref{111g} through the numerical interpolation approach.

\begin{figure}[!ht]
\caption{Massive modes from a thick brane canonical and non-canonical.}  \label{fig7}
  \centering
  \subfigure[Canonical case.]{\includegraphics[height=7cm,width=8cm]{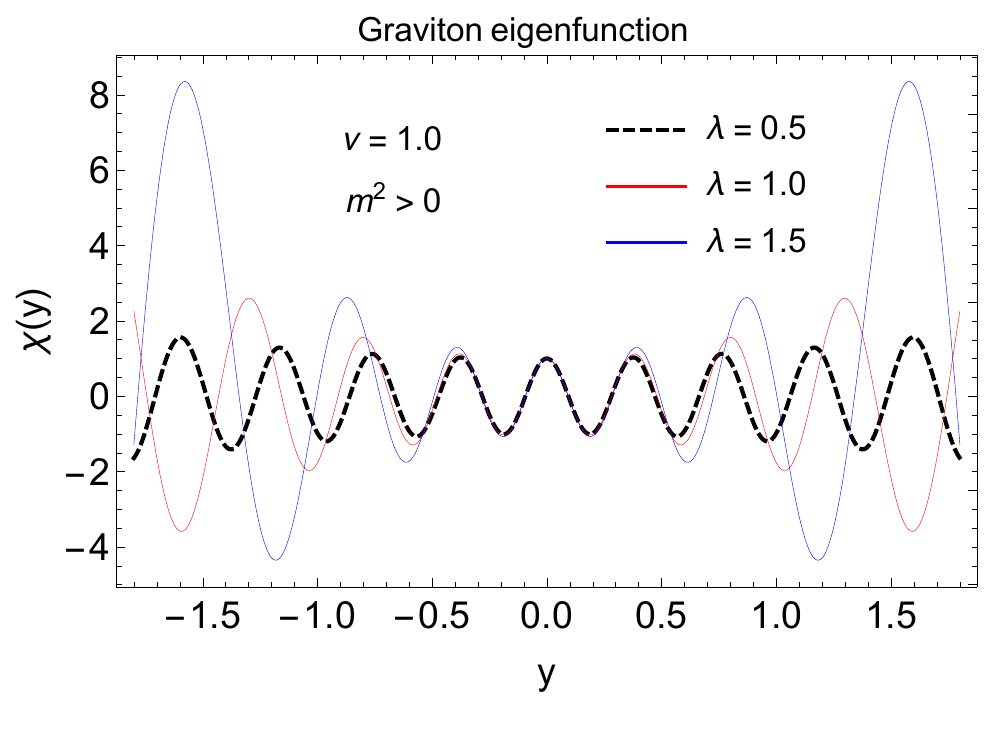}}\hfill
  \subfigure[Non-canonical case.]{\includegraphics[height=7cm,width=8cm]{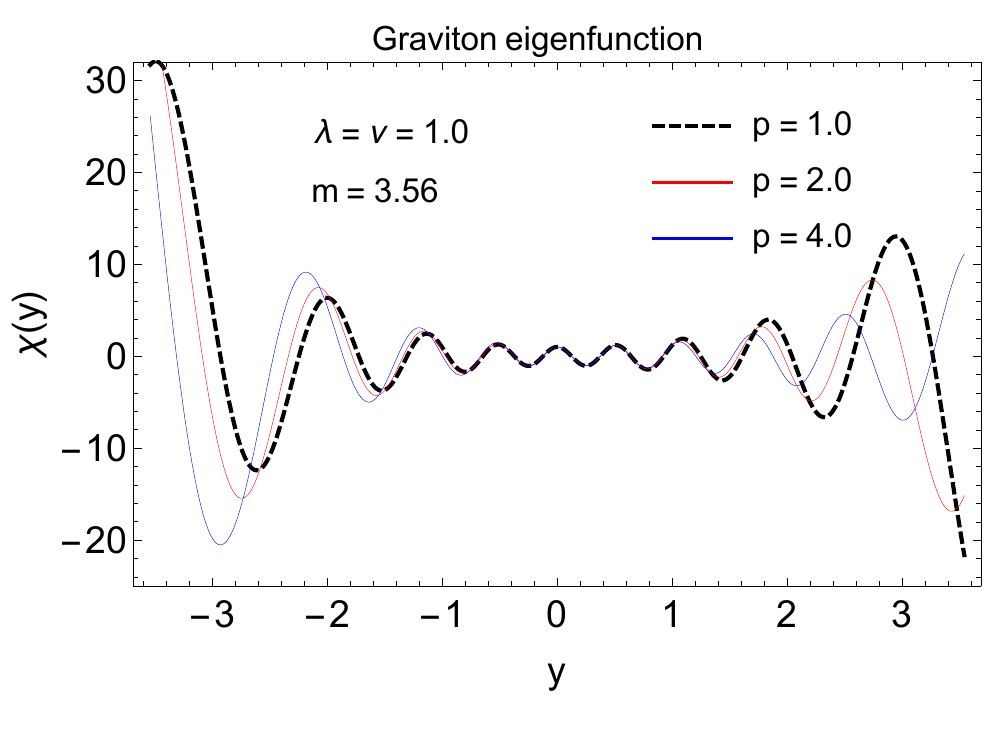}}
\end{figure}

For both cases studied so far, one notes that the parameters $\lambda$ (associated with the brane's width) and $p$ (responsible for the brane asymmetry) increase the amplitude of the KK modes [Fig. \ref{fig7}]. However, near the brane, one obtains similar oscillation amplitudes. Meanwhile, in the canonical case, far away from the brane, the KK modes increase the amplitude while maintaining oscillations symmetric concerning the brane, i.e., $y=0$. In contrast, in the non-canonical conjecture, far away from the brane, the amplitudes of the KK modes increase asymmetrically. Thus, the results of the zero modes [Fig. \ref{fig6}], along with the results of the KK modes, ensure the absence of the brane splitting phenomenon.

\subsubsection{The resonant modes}

A topic of meaningful interest within the framework of braneworlds is the analysis of resonant modes in the theory. The study of resonances is crucial because it provides valuable insights into the massive modes, allowing for an understanding of the coupling between these modes and the matter confined on the brane \cite{CASA1,Liu1,Liu2,BazeiaX}. To conduct this analysis, one needs to determine the amplitudes of the wave function $\psi_m (z)$ on the brane. Hence, to accomplish our purpose, we examine the solutions $\psi_m (z)$ of the Eq. \eqref{RevEq1} normalized at $z=0$. As exposed in Refs. \cite{Cruz,Csaki}, modes with a mass significantly greater than $V_{\textrm{eff}}^{(\textrm{max})}$, the effective potential introduces only small perturbations. Therefore, modes of the wave function $\psi(z)$ for which $m^2<V_{\textrm{eff}}^{(\textrm{max})}$ may resonate with the potential. Thus, the contribution $\vert\xi\psi_m(0)\vert^2$ provides us the probability of locating the massives modes at $z=0$\footnote{\revision{In this framework, $\xi$ is the normalization constant of the mode $\psi_m(0)$.}} \cite{Cruz}. One defines the relative probability of the resonant modes as 
\begin{align}\label{RevEq1}
    P(m)=\frac{\vert\psi_m(0)\vert^2}{\int_{-100}^{100}\vert\psi_m(z)\vert^2\,dz}.
\end{align}

\begin{figure}[!ht]
    \centering
  \includegraphics[height=7cm,width=8cm]{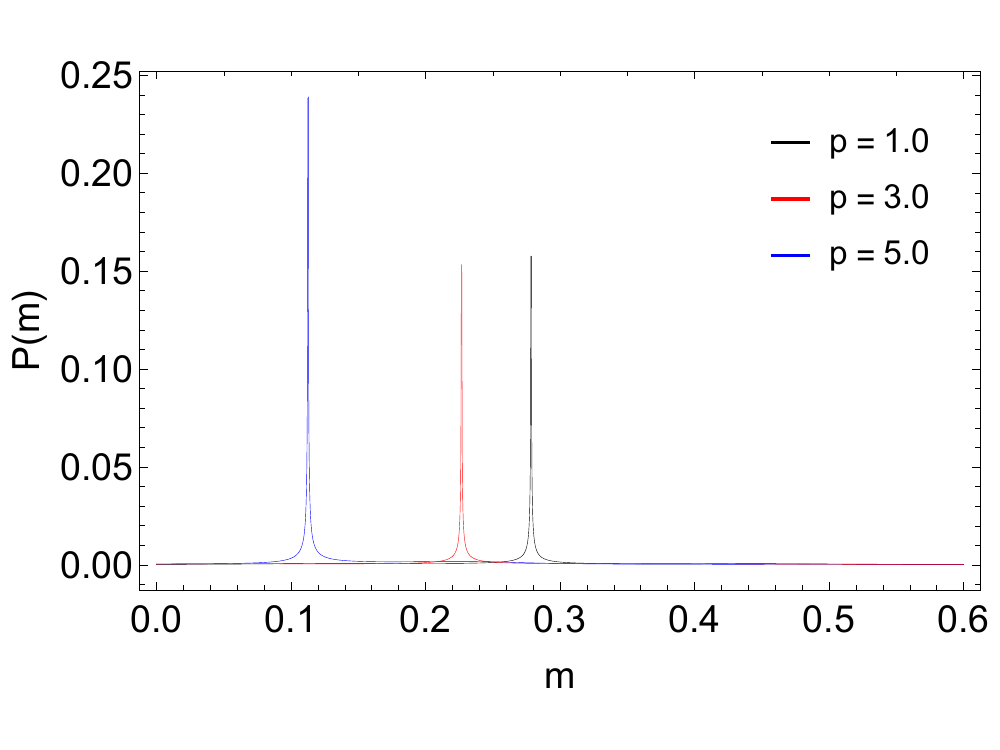}\vspace{-1cm}
  \caption{The relative probability $P(m)$ of the resonant modes when $\lambda=\nu=1$.}  \label{Revfig1}
\end{figure}

In Fig. \ref{Revfig1}, we present the relative probability associated with the resonant modes within the non-canonical braneworld scenario\footnote{\revision{The canonical scenario leads to results resembling those obtained in the non-canonical case. Therefore, to avoid ambiguities, we restrict our analysis to the influence of the non-canonical scenario on the resonance phenomenon.}}, as the parameter $p$, which is responsible for the asymmetry of the brane, is varied.

By inspection of the numerical results, one infers that for $p=1,2,$ and $3$, the resonance peaks occur, respectively, at $m=0.278\pm 10^{-3}, 0.226\pm 10^{-3}$, and $0.112\pm 10^{-3}$. These results allow us to deduce that the probability of light or massless modes coupled to the brane is higher than that of heavier modes. Moreover, the results indicate that when the asymmetry of the five-dimensional braneworld increases, the value of the parameter $m$ concerning resonant light modes decreases. Besides, one highlights that the resonant structures tend to disappear in agreement with results announced by the localization of zero modes.

\subsection{Scalar perturbations}

To conclude the study on the braneworld stability, let us proceed to the analysis of scalar perturbations. To fulfill our purpose, we will adopt metric perturbation
\begin{align}
    ds^2=\e^{2A(z)}\bigg[(1+2\Psi(x,z))\eta_{\mu\nu}dx^\mu dx^\nu+(1+2\Phi(x,z))dz^2\bigg].
\end{align}
Furthermore, one considers small scalar field perturbation in the form $\phi(y)\rightarrow\overline{\phi}(y)+\epsilon(x^\mu,y)$.
Thus, the perturbed Einstein equation turns into
\begin{align}
\label{ps1}
\eta_{\mu\nu}\bigg(2\square\Psi+2\Ddot{\Psi}+14\dot{A}\dot{\Psi}-4\Phi\Ddot{A}-12\Phi\dot{A}^2-2\dot{A}\dot{\Phi}\bigg)\nonumber\\+2\partial_\mu\partial_\nu (\Phi+2\Psi)=-\frac{4}{3}\eta_{\mu\nu}\textrm{e}^{2A}\frac{\partial V}{\partial\phi}\epsilon-\frac{8}{3}V\Psi\eta_{\mu\nu}\textrm{e}^{2A},
\end{align}
\begin{align}
\label{ps2}
2\square\Phi+8\Ddot{\Psi}-8\dot{A}\dot{\Phi}+8\dot{A}\dot{\Psi}-2\phi(2\Ddot{A}+\dot{A}^2)=\nonumber\\-2f_{\phi}\epsilon\dot{\phi}^2-4f\dot{\phi}\dot{\epsilon}-\frac{4}{3}\e^{2A}\frac{\partial V}{\partial\phi}\epsilon-\frac{8}{3}V\Phi\textrm{e}^{2A},
\end{align}
and
\begin{align}
\label{ps3}
&-2\partial_\mu\dot{\Psi}+2\dot{A}\partial_\mu\Phi=\frac{2}{3}f\dot{\phi}\,\partial_\mu\epsilon.
\end{align}

Concurrently, the perturbed matter equation takes the form
\begin{align}
f\bigg[\square\epsilon+\Ddot{\epsilon}+3\dot{A}\dot{\epsilon}+2\dot{\phi}\bigg(2\dot{\Psi}-\frac{1}{2}\dot{\Phi}\bigg)-2\Phi(\Ddot{\phi}+3\dot{A}\dot{\Phi})\bigg]\nonumber\\+f_{\phi}\,\epsilon(\Ddot{\phi}+3\dot{A}\dot{\phi})-2\Phi f_{\phi}\dot{\phi}^2+f_{\phi\phi}\dot{\phi}\,\epsilon+2f_{\phi}\dot{\phi}\,\dot{\epsilon}=\e^{2A}\frac{\partial^2 V}{\partial\phi^2}\epsilon.   
\end{align}

By inspection of the Eqs. \eqref{ps1} and \eqref{ps3}, one notes the presence of off-diagonal contributions in the geometric sector. Naturally, these contributions lead us to two constraints between $\Phi$, $\Psi$, and $\epsilon$, viz.,
\begin{align}
\partial_\mu\partial_\nu(\Phi+2\Psi)=0,
\end{align}
and
\begin{align}
2f\dot{\phi}\,\epsilon+6(\dot{\Psi}+\Psi\dot{A})=0.
\end{align}
These constraints lead us to a single physical scalar degree of freedom, namely, 
\begin{align}
\square\Psi+\Ddot{\Psi}+\bigg(3\dot{A}-\frac{2\Ddot{\phi}}{\dot{\phi}}\bigg)\dot{\Psi}+\bigg(4\Ddot{\Psi}-\frac{4\dot{A}\Ddot{\phi}}{\dot{\phi}}\bigg)\Psi=0.    
\end{align}

In search of a perturbed Schrödinger-like equation, we apply the variable change $\Psi^{\ast}\to\e^{-\frac{3}{2}A}\frac{\Psi}{\dot{\phi}}$, which leads us to
\begin{align}\label{spe}
-\ddot{\Psi}^{\ast}+V_{\textrm{s}}\,\Psi^{\ast}=m_{\textrm{s}}^2\,\Psi^{\ast},
\end{align}   
where the scalar effective potential is
\begin{align}\label{vv1}
V_{\textrm{s}}(z)=-\frac{5}{2}\Ddot{A}+\frac{9}{4}\dot{A}^2-\frac{\dddot{\phi}}{\dot{\phi}}+\dot{A}\frac{\Ddot{\phi}}{\dot{\phi}}+2\bigg(\frac{\Ddot{\phi}}{\dot{\phi}}\bigg)^2.    
\end{align}

\begin{figure}[!ht]
\caption{\revision{Scalar effective potential vs. extra-dimensional coordinate.}}  \label{Revfig2}
  \centering
  \subfigure[\revision{Canonical case.}]{\includegraphics[height=7cm,width=8cm]{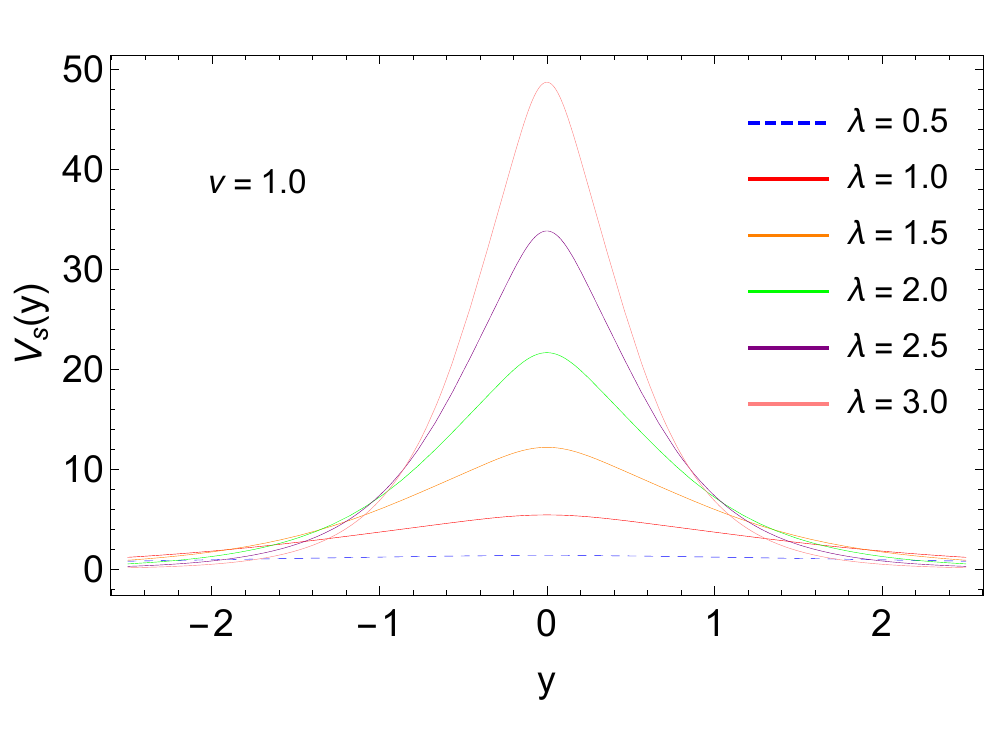}}\hfill
  \subfigure[\revision{Non-canonical case.}]{\includegraphics[height=7cm,width=8cm]{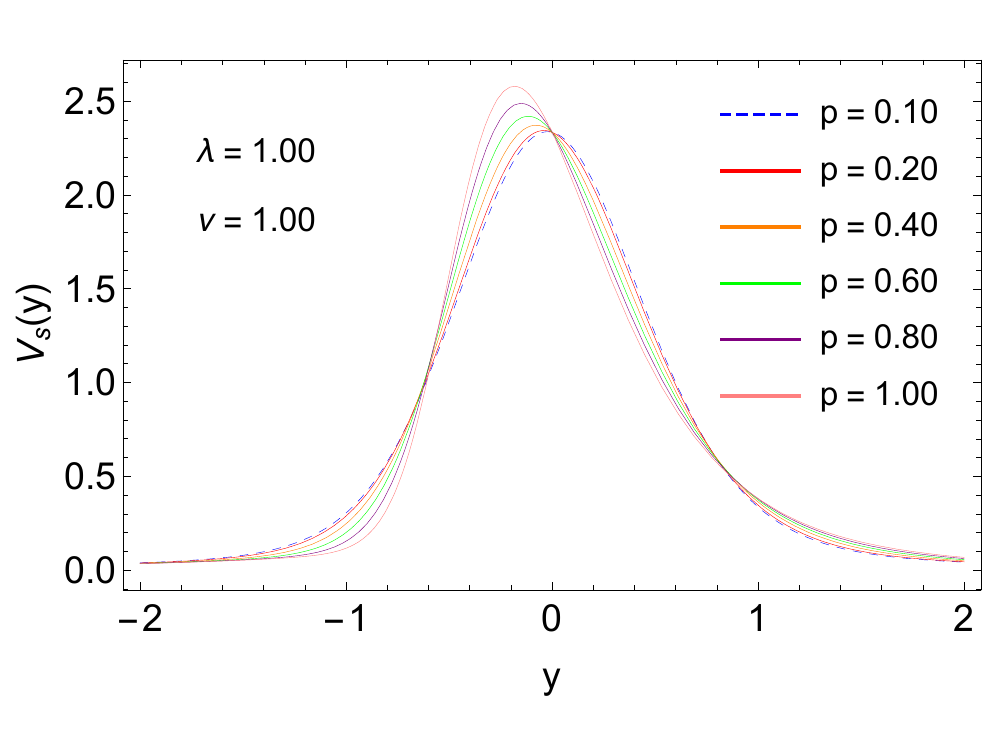}}
\end{figure}

Once again, the dot notation is the derivative concerning the conformal variable $z$. Therefore, we reformulate the Eq. \eqref{vv1} to an expression regarding the extra-dimensional variable $y$ boil down to
\begin{align}\nonumber
V_{\textrm{s}}(y)=&-\e^{2A}\bigg[\frac{1}{4}A^{\prime\,2}+\frac{5}{2}A^{\prime\prime}+\frac{\phi^{\prime\prime\prime}+3A^{\prime}\phi^{\prime\prime}+2A^{\prime\,2}\phi^{\prime}+A^{\prime\prime}\phi^{\prime}}{\phi^{\prime}}-A^{\prime}\bigg(\frac{A^{\prime}\phi^{\prime}+\phi^{\prime\prime}}{\phi^{\prime}}\bigg)+\\
-&2\bigg(\frac{A^{\prime}\phi^{\prime}+\phi^{\prime\prime}}{\phi^{\prime}}\bigg)^2\bigg]   
\end{align}
 We show the scalar perturbed potential ($V_s$) in Fig. \ref{Revfig2}.

By analyzing the potential $V_s(y)$ for the two models, i.e., the canonical and non-canonical models [see Figs.
\ref{Revfig2}(a) and \ref{Revfig2}(b)], one can note that, in the canonical theory, there is a potential barrier localized on the brane ($y=0$) whose amplitude symmetrically increases as the parameter $\lambda$ grows. In contrast, in the non-canonical case, as the parameter $p$ (originating from the generalizing function) increases, the scalar perturbed potential also increases in amplitude but becomes asymmetric concerning the brane. Thus, the profile of barrier-like potential in both cases ensures non-localized scalar modes.


\section{Final Remarks}\label{secIV}

In this study, we investigated a five-dimensional non-canonical braneworld theory. Besides, we studied the emergence of thick domain walls on the $\textrm{AdS}$ space. Analyzing the projection of four-dimensional fluctuations allowed us to infer the behavior of the graviton. Additionally, we announced the existence of asymmetric brane by checking the matter field profile, energy, and KK spectrum.

Considering our models, one can note that in the canonical case, altering the $\lambda$-parameter results in a modification in the brane's width. Furthermore, we concluded that spontaneous symmetry breaking is responsible for the emergence of thick domain walls. Indeed, this is a consequence of the obtained warp function. Besides, one noted that the domain walls are localized kinks on the 3-brane.

In the non-canonical case, the warped geometry of the thick brane becomes asymmetric when the $p$-parameter varies. Thus, we noted that the asymmetry of spacetime affects the matter sector, resulting in a vacuum decay. This decay induces an asymmetry in the matter field, deforming the domain wall, making the matter field emitted by the 3-brane more compact at $y \in (-\infty, 0]$. Therefore, this gives rise to an asymmetric thick brane.

Furthermore, we concluded that the localized four-dimensional gravity requires zero modes normalizable, i.e.,
\begin{align}
    \int_{-\infty}^{\infty}\,dy\,\textrm{e}^{2A(y)}<\infty. 
\end{align}
which leads us to
\begin{align}
    \lim_{y\to\pm\infty}\textrm{e}^{2A(y)}=0.
\end{align}
Thus, the massive modes indicate that, in both cases, changes in the brane's width or the p-parameter (parameter of the generalizing function) modify the amplitudes of the KK modes. Meanwhile, near the brane, the oscillation amplitudes remain approximately constant. In the canonical case, as one moves away from the brane, the amplitudes of the KK modes increase while maintaining symmetric oscillations around $y=0$. In contrast, in the non-canonical case, far from the brane, the amplitudes of the KK modes increase asymmetrically.

Upon analyzing the resonance phenomenon, one noted that as the brane's asymmetry increases, the mass associated with the resonance peaks decreases. This inspection allows us to conclude that the probability of the light or massless modes coupled to the brane is higher than that of heavier modes. Additionally, the resonant framework tends to disappear according to the location of the zero modes. Finally, the results concerning the scalar stability of the theory reveal the presence of symmetric and asymmetric potential barriers, ensuring that the scalar modes in both theories are non-localizable.


\section{Acknowledgment}
The authors wish to express their gratitude to FAPEMA and CNPq (Brazilian research agencies) for their invaluable financial support. F. C. E. Lima is supported by FAPEMA BPD-05892/23. F. M. Belchior is supported by CNPq 161092/2021-7. C. A. S. A. is supported by CNPq 309553/2021-0 (CNPq/Produtividade) and project 200387/2023-5. Furthermore, C. A. S. A. acknowledges the Department of Theoretical Physics \& IFIC from the University of Valencia for their warm hospitality.  P. K. S. acknowledges Science and Engineering Research Board, Department of Science and Technology, Government of India for financial support to carry out Research project No.: CRG/2022/001847.

\section{Conflicts of Interest/Competing Interest}

The authors declared that there is no conflict of interest in this manuscript.


\end{document}